\documentclass[conference]{IEEEtran}

\usepackage{times}

\usepackage{graphicx}
\usepackage{epsfig}
\usepackage{subfigure}
\usepackage[cmex10]{amsmath}
\usepackage{epstopdf}
\usepackage{multirow}
\usepackage{array}
\usepackage{cite}
     \usepackage{balance}

\begin{document}
%
\title{ Level-Shifted Neural Encoded Analog-to-Digital Converter}
\author{\IEEEauthorblockN{Aigerim Tankimanova, Akshay Kumar Maan, and Alex Pappachen James}
\IEEEauthorblockA{
School of Engineering, Nazarbayev University, Astana\\
www.biomicrosystems.info/alex\\
Email: atankimanova@nu.edu.kz; apj@ieee.org}}


\maketitle

\begin{abstract}

This paper presents the new approach in implementation of analog-to-digital converter (ADC) that is based on Hopfield neural-network architecture. Hopfield neural ADC (NADC) is a type of recurrent neural network that is effective in solving simple optimization problems, such as analog-to-digital conversion. The main idea behind the proposed design is to use multiple 2-bit Hopfield NADCs operating as quantizers in parallel, where analog input signal to each successive 2-bit Hopfield ADC block is passed through a voltage level shifter. 
This is followed by a neural network encoder to remove the quantization errors.  In traditional Hopfield NADC based designs, increasing the number of bits could require proper scaling of the network parameters, in particular digital output operating region.  Furthermore, the resolution improvement of traditional Hopfield NADC creates digital error that increases with the increasing number of bits. The proposed design is scalable in number of bits and number of quantization levels, and can maintain the magnitude of digital output code within a manageable operating voltage range.  

\end{abstract}
\section{Introduction}

Neurons have an inherent ability to convert analog input signals to digital signals through the firing synaptic functions in axons and learn input responses through weighted input addition operations in its dendrites.
The architecture of Neural Analog-to-Digital Converters (NADC) is inspired from the firing and learning mechanism of neural networks \cite{tank1986simple, guo2015modeling,wang2014adaptive}.  Hopfield neural network is known to be one of the prominent methods used to build  the NADC, where the analog-to-digital (ADC) task is introduced as a simple optimization problem \cite{tank1986simple}. The Hopfield network is a type of simple recurrent neural network that uses one layer of analog processing units (neurons). The outputs of this layer are fed back to the network through a predetermined set of weights (synapses). The system is characterized by a computational cost function (or energy function, $\displaystyle\it{E}$) that tends to reach the minimum value during the computational process. The  minimum values of energy function are stored in the network by preprogramming the synapses such that at certain input the system energy function will reach minimum \cite{80300}. This system deduces the corresponding solution using parallel processing of input information by neurons that indicates the property of collective computation of highly interconnected neural networks \cite{hopfield1982neural}.

Ideally the network should provide one particular solution for the corresponding input at the global minima state of energy function, however, in practice the Hopfield network provides more than one solution for each input level \cite{tank1986simple}. In other words, the energy function of the Hopfield network ADC has additional local minima states for each input \cite{tank1986simple}.  The local minima behavior introduces digital errors at the output. This problem occurs due to the changing neurons' thresholds during the computation process. Different methods were proposed to mitigate this problem, such as periodically resetting the neurons to initial state or initial threshold value, thereby, minimizing the hysteresis neuron response \cite{tank1986simple}. Another method of eliminating local minima states was proposed by Lee and Sheu \cite{80300} is applying correction currents to the input of neurons in order to fill up undesired local minima states. In this work we reduced the number of bits in a single Hopfield ADC block to 2-bit size in order to minimize digital error from local minima states. Therefore, the proposed method can be applied to small analog input signal ranges that require higher resolution.

The highly interconnected architecture of Hopfield NADC requires appropriate scaling in order to obtain higher resolution. Thus, the levels of output voltages should be high enough for proper operation of the network and at the same time they must comply with the digital circuitry that would follow the ADC. Moreover, by increasing number of bits the degree of digital uncertainty also increases due to multiple energy function minima states that are created during conversion \cite{tank1986simple}. Therefore, Hopfield NADC is not widely applied in real systems due to such impracticality. This paper presents the new two-stage ADC design built with multiple 2-bit Hopfield NADCs and artificial neural network encoder operating in parallel. The input analog signal to each successive 2-bit ADC is DC shifted by a constant voltage and forms the quantization stage. The neural network encoder reduces the quantization errors to ensure robust ADC function. The advantage of the proposed idea is that it can be applied in low power systems that require good resolution, for instance, pixel-parallel ADC.

In Section II of the current paper the background on the of the proposed idea and the theory of original Hopfield neural network ADC is presented. Section III describes details of the proposed ADC design. In the Section IV the simulation results for 16  quantization levels is shown. Section V presents the discussion of the introduced work.




\section{Background}

The understanding of how the mechanisms of human brain are governed is one of the modern challenges in science and engineering. One of the earliest works in this field was presented by McCulloch and Pitts\cite{mcculloch1943logical}, where the authors proposed a mathematical model of an artificial neural network (ANN) that performs logic operations and is built with binary neurons. Based on that binary neuron model, Hopfield developed a network that can solve computational problems and also can act as a simple associative memory network\cite{hopfield1982neural}. In contrast to feed-forward ANN architectures, such as Perceptron\cite{widrow1960adaptive}, Hopfield Neural Network consists of a single neuron layer with feedback connections. 
The  feedback network structure is inspired from the feedback properties of biological neural networks.\cite{hopfield1986collective}


The hardware implementation of a 4-bit neural ADC proposed by Hopfield is shown on Fig. 1. Each node represents a synapse that is characterized by a conductance value, $\displaystyle \it{T}$. 
\begin{figure}[htp]
\centering
\includegraphics[width=70mm]{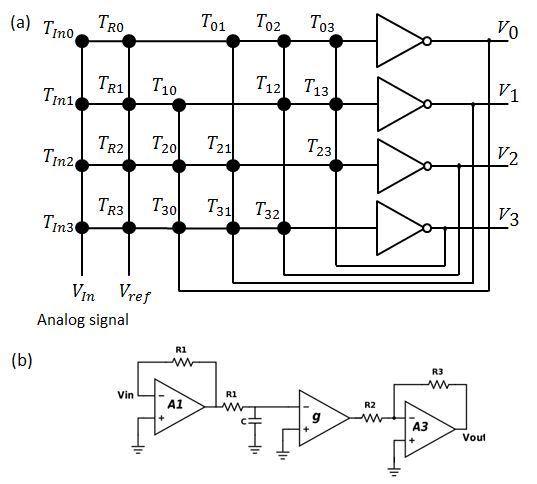}
\caption{(a) 4-bit Hopfield neural network ADC \cite{tank1986simple}, (b) neuron implementation.}
\label{f1}
\end{figure}
Each neuron performs a weighted summation (through synapses) of input currents from the input signal and outputs from other neurons \cite{42986}. The output of $\displaystyle \it{i-}$th neuron can be described by the function $\displaystyle \it{V_i = g_i(u_i)}$, where $\displaystyle \it{u_i}$ is the input voltage to the neuron. The activation function $\displaystyle \it{g_i(u_i)}$ is  characterized by monotonic sigmoid response: 
\begin{equation}
g_i(u_i)= \frac{1}{1+exp(u_i)}
\label{Eq1}
\end{equation}
The output of neuron can be of two values $\displaystyle \it{V_i}$ = 1 (logic high) if the input is higher than the neuron threshold, and $\displaystyle \it{V_i}$ = 0 (logic low) if the input is lower than the threshold \cite{hopfield1984neurons}. The desired neuron response is obtained by using op-amp comparator as shown in Fig. 1(b). Resistors are used to represent interconnection weights and the conductances. 

The important conditions \cite{tank1986simple} that satisfy the network convergence to the local minima are: (1) zero diagonal elements $\displaystyle \it{T_{ii}}$ = 0, so that neuron will not feed the  output back to its input, and (2) the symmetry condition $\displaystyle \it{T_{ij}}$ = $\displaystyle \it{T_{ji}}$ \cite{hopfield1984neurons}. Another valuable condition is that the neuron output voltage value should  be related to analog input voltage as:
\begin{equation}
{V_{In}} = \sum_{i=0}^{N-1}2^{i}\it{V_{i}}
\label{Eq2}
\end{equation}
Under these conditions, the system should provide the correct solutions for corresponding input voltage levels.


Assuming neurons are of infinite input and zero output resistances the current flowing into the $ \displaystyle \it{i-}$th neuron can be described by the following Eq. \ref{Eq2}: 
\begin{equation}
C \frac{du_i}{dt} = \sum_{j}T_{ij}V_j-(T_{Ini} + T_{Ri} + \sum_{j}T_{ij} )u_i+ T_{Ini}V_{In}+T_{Ri}V_{ref}
\label{Eq3}
\end{equation}
where $\displaystyle\it{C\frac{du_i}{dt}}$ is current flowing into the neuron, $\displaystyle \it{V_j}$ is the output from neuron $\displaystyle \it{j}$ that is connected to the neuron $\displaystyle \it{i}$ through a conductance $\displaystyle \it{T_{ij}}$. The analog input signal is represented by $\displaystyle \it{V_{In}}$ and it is connected to the neuron through the conductance $\displaystyle \it{T_{Ini}}$. The reference voltage $\displaystyle \it{V_{ref}}$ is connected to the neuron input through the conductance $\displaystyle \it{T_{Ri}}$. The term $\displaystyle \it{T_{Ini} + T_{Ri} + \sum_{j}T_{ij}}$ represents the neuron input effective conductance \cite{guo2015modeling}. 

The system is described in more general using the energy function Eq. \ref{Eq4} for N number of processing units \cite{hopfield1984neurons}. 
\begin{equation}
E=-\frac{1}{2}\sum_{i=0 j=0}^{N-1}T_{ij}V_iV_j-\sum_{i=0}^{N-1}V_iI_i+\sum_{i=0}^{N-1}T_i\int_{0}^{V_i}g^{-1}(V)dV
\label{Eq4}
\end{equation}
Where $\displaystyle \it{I_{i}}$ represents analog input current flowing into neuron $\displaystyle \it{i}$ and $\displaystyle \it{T_{i}}$ is input effective conductance of $\displaystyle \it{i-}$th neuron. 
Using the Eqs. \ref{Eq3} and \ref{Eq4} the conductance values can be calculated as \cite{tank1986simple}:
\begin{equation}
T_{ij}=2^{i+j}, T_{Ini}=2^{i}, T_{Ri}=2^{2i-1}
\label{Eq5}
\end{equation}
Using the Eqs. \ref{Eq1}-\ref{Eq5} it is possible to construct a circuit for Hopfield NADC of any size. However,  the behavior of circuit elements introduces uncertainties that corrupts the digital output of the ADC limiting its scalability \cite{tank1986simple}. 

\section{Proposed Design and Results}

Fig. 2 represents the block diagram of the proposed idea. The diagram consists of 2 connected blocks, namely, a quatizer and an encoder. The quantizer block consists of a series of 2-bit Hopfield ADCs, each acting as the basic unit to quantize the input signal. The input signal is concurrently applied to the $\displaystyle\it{n}$ number of separate 2-bit neural ADC quantizers through ($\displaystyle\it{n-}$1) analog voltage level shifters. Therefore, each successive 2-bit ADC quantizer receives the analog signal that is level-shifted to  $\Delta\displaystyle\it{V}$  relative to the previous ADC input signal. The output of the Hopfield level shifted ADC is applied to the encoder block where a three layer artificial neural network that uses back-propagation algorithm to encode the binary pattern to weighted binary pattern. The errors from the quantizer stage are compensated by a well trained neural network at the encoder stage.  
\begin{figure}[ht]
\centering
\includegraphics[width=90mm]{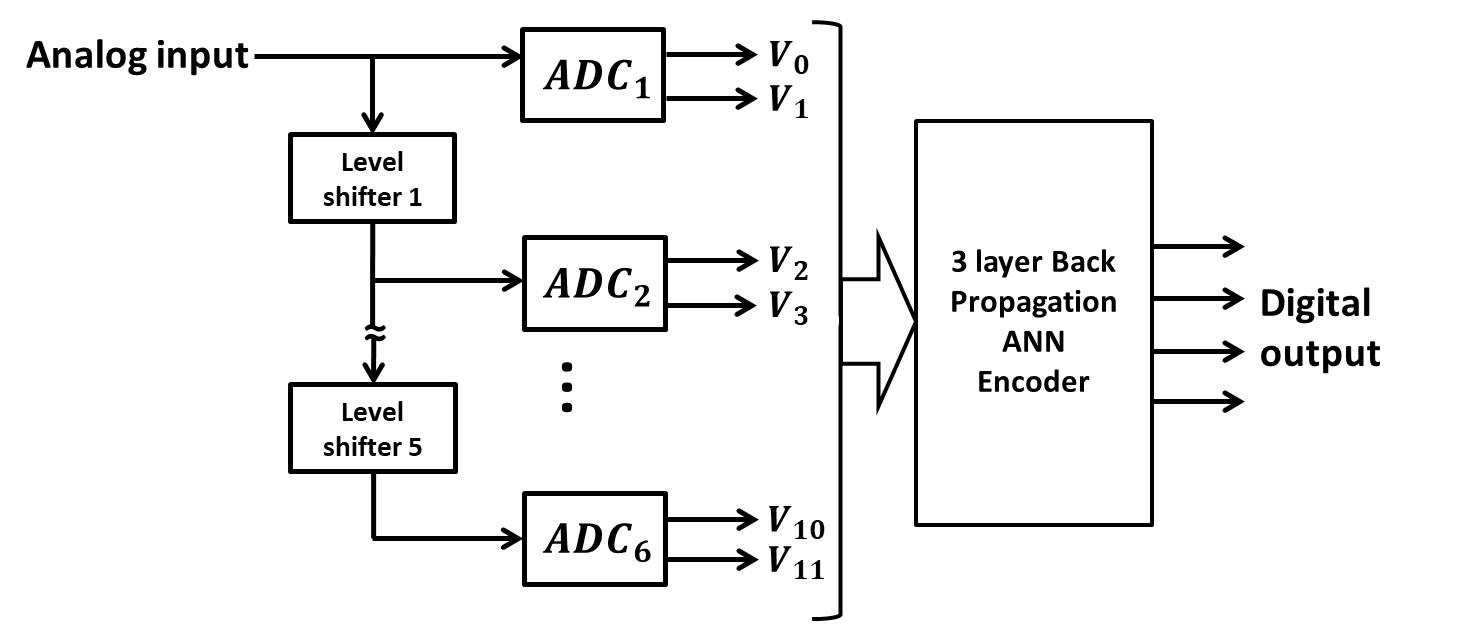}
\caption{ The block diagram of the proposed design. Quantizer and encoder blocks.}
\label{f2}
\end{figure}

For the accurate operation of the ADC, the analog input voltage range and the neuron maximum output must satisfy the condition in Eq. \ref{Eq2}. In the proposed design, the analog input voltage ${V_{In}}$ is kept within the range [0, 2]V. The corresponding neuron output voltage value ${V_{i}}$ is calculated to be at -670 mV. The negative sign shows the inverting property of neurons suggested by Guo et al.\cite{guo2015modeling} to keep conductance positive. The values of conductance must match circuit parameters such as the input impedance of op-amps. In addition, the conductance should be scaled  to comply with the condition in Eq. \ref{Eq2}. Therefore, the conductance values can be calculated by modifying Eq. \ref{Eq5} to the following form, as suggested by: \cite{tank1986simple}.
\begin{equation}
T_{ij}=\frac{2^{i+j}}{V_i}, T_{Ini}=\frac{2^{N+i}}{V_{In}}, T_{Ri}=2^{i-1}+\frac{2^{2i-1}}{V_{ref}}
\label{Eq6}
\end{equation}
where the $\displaystyle \it{V_{In}}$ is the maximum input voltage set to 2 V and $\displaystyle\it{V_{ref}}$ is set to -670 mV. 
\begin{table}[]
\setlength\extrarowheight{5pt}
\centering
\caption{Calculated parameters}
\label{my-label}

\begin{tabular}{|c|c|c|c|l}
\cline{1-4}
Conductance         & Value ($\mu$S) & \begin{tabular}[c]{@{}c@{}}Neuron \\     resistances\end{tabular} & Value (kOhm)         &  \\ \cline{1-4}
$\it{T}_{01}$ & 29.9  & \multirow{2}{*}{$\it{R}$1}                                      & \multirow{2}{*}{1}   &  \\ \cline{1-2}
$\it{T_{10}} $ & 29.9  &                                                                   &                      &  \\ \cline{1-4}
$\it{T_{R0}} $ & 12.5  & \multirow{2}{*}{$\it{R}$2}                                      & \multirow{2}{*}{100} &  \\ \cline{1-2}
$\it{T_{R1}}$  & 39.9  &                                                                   &                      &  \\ \cline{1-4}
$\it{T_{In0}}$ & 20.0  & \multirow{2}{*}{$\it{R}$3}                                      & \multirow{2}{*}{2}   &  \\ \cline{1-2}
$\it{T_{In1}}$ & 40.0  &                                                                   &                      &  \\ \cline{1-4}
\end{tabular}

\end{table}

Table 1 shows the conductance values and the neuron resistances used in simulations. Using the proposed method we obtain 16 and quantization level ADC with 6 blocks of 2-bit Hopfield ADC quantizers. In the proposed ADC, the analog input is shifted 5 times, i.e. 0.1V after each ADC block. 


The constructed 2-bit Hopfield ADC quantizer is simulated by applying sinusoidal and linear inputs as shown in Fig. 3. The reference voltage is set to -670 mV. However, for multiple 2-bit ADC blocks with level shifters the reference voltages affects the digital outputs. With adjustment of reference voltages of each ADC block, quantization of the analog input can be obtained. Fig. 4(a) represents the Integral Nonlinearity (INL) analysis for the obtained transfer characteristics of 16 levels ADC shown on Fig. 4(b). As it can be observed, the maximum INL is about 1 LSB (least significant bit). The adjustment of reference voltages was performed manually and the resultant transfer characteristics exhibit quantization error. The accuracy can be improved by using a neural network encoder that will perform the error correction and encoding of output from level-shifted NADC. It should be noted that these results are shown for Hopfield ADC without the neural encoder. The simulations with a three layer neural network using back-propagation with learning rate of 0.3 and momentum of 0.9, 12 neurons in first layer, 11 neurons in hidden layer and 4 neurons in output layer  reduces the errors to zero.

\begin{figure}[ht]
\centering
\includegraphics[width=90mm]{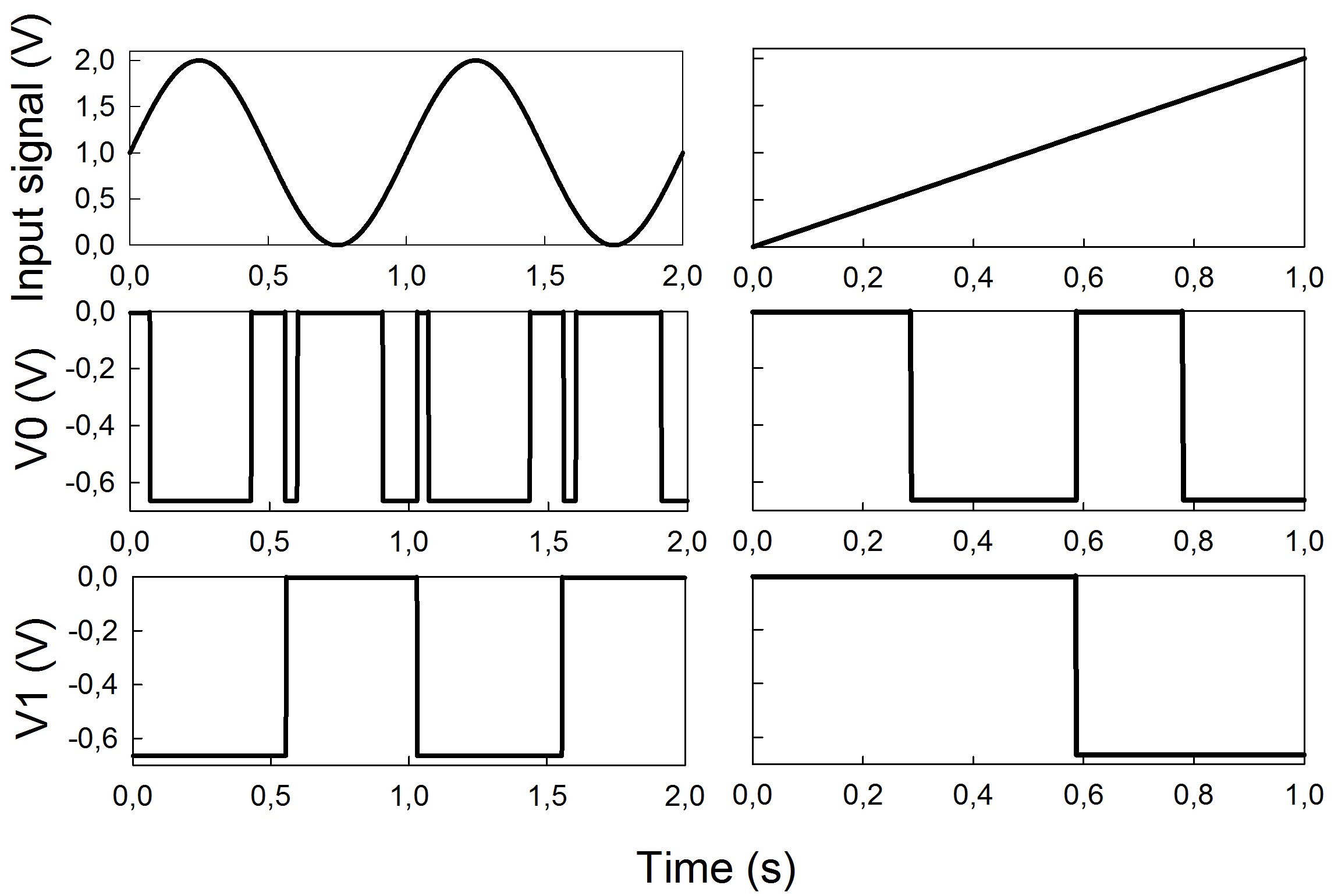}
\caption{2-bit Hopfield ADC simulation results}
\label{f5}
\end{figure}
\begin{figure}[ht]
\centering
\includegraphics[width=80mm]{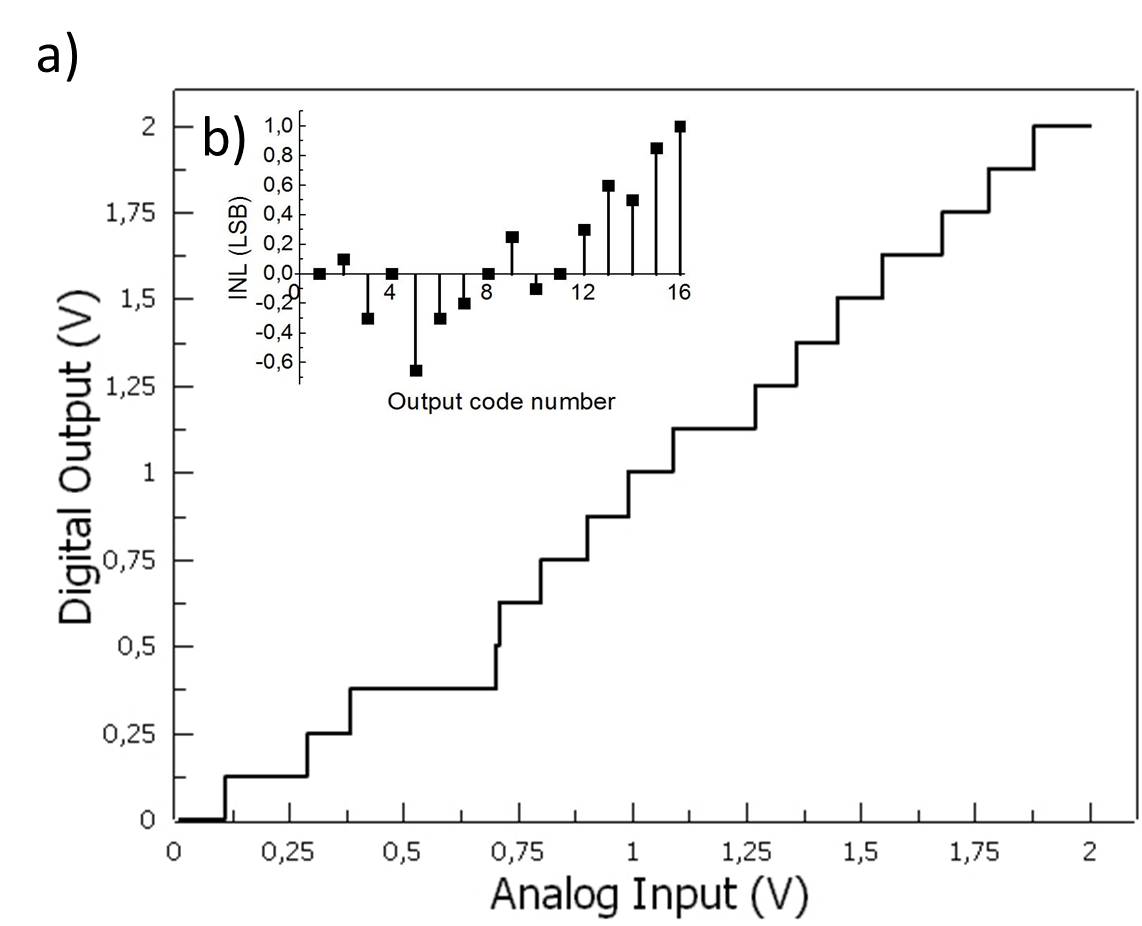}
\caption{(a)  Transfer characteristics for 16 quantization levels level-shifted ADC,(b) The INL analysis of the proposed ADC design.}
\label{f6}
\end{figure}

\begin{figure}[ht]
\centering
\includegraphics[width=80mm]{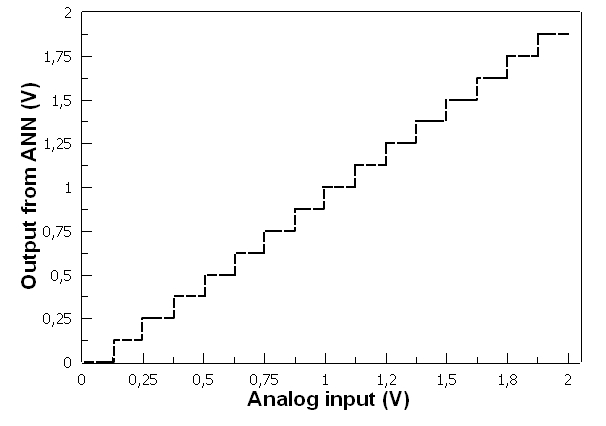}
\caption{Transfer characteristics of the proposed ADC after neural encoder correction.}
\label{f7}
\end{figure}

\section{Discussion and Conclusion}
From the transfer characteristics of 16 quantization levels ADC the positive gain error is observed with the maximum of 1 LSB, and using the neural network encoder the bit error reduces to 0. As it was already mentioned, the reference voltages at each ADC block affect the digital output signal of the ADC. Reference voltage is applied at each ADC block separately. The memory stored in the network is affected by reference voltage value, as it alters the location of energy function minima. It is observed, that with higher magnitude of reference voltage the conversion in 2 bit Hopfield ADC quantizer starts at higher values of analog input so that the dynamic range of the converter is altered. Further work is to be conducted to reduce the value of quantization error of the current design. The output codes of obtained ADC configuration do not support conventional 4 bit format. An additional neural network encoder part solves the quantization error problem occurring in the proposed design. 

The original Hopfield neural network design was developed in 1980s and the idea was popular during 1990s. Since that period there is a number of interpretations and different versions of such architecture. However, this design was not practically used and is still an exploratory topic of study. Nevertheless, the simplicity of Hopfield neural network and its crossbar architecture has attracted modern scientists in \cite{guo2015modeling}, \cite{wang2014adaptive} where they proposed hybrid CMOS/memristor based architecture. The discovery of memristive device that can mimic the behavior of biological synapses \cite{strukov2008missing} attracted many scientists as it opened possibility of implementing low-power consumption neural networks in hardware \cite{liu2015spiking}, \cite{jo2010nanoscale}. Therefore, the addition of memristor to the current design is considered to be reasonable because it is possible to reduce on-chip area and power consumption when memristor is used as synaptic interconnect. It is planned to work in the direction of hybrid CMOS/memristor design based level-shifted ADC. 
 
In this paper, we presented a multiple 2-bit Hopfield neural ADC quantizers with level shifter circuits placed between each ADC block. In contrast to traditional Hopfield ADCs where the increase in the number of bits results in increase in bit errors, the proposed architecture does not have bit errors.
In the traditional Hopfield ADC design to obtain higher number of bits, the analog input and output voltages should be scaled for the proper operation of the circuit. In this work, we have demonstrated that the high resolution ADC can be achieved with  small voltage range without the issues of scaling the voltages and  conductance. The proposed design can be applied in such small voltage systems, as pixel-parallel ADCs where a good resolution is a big advantage.



\bibliographystyle{IEEEtran}
\bibliography{References}
 \balance
\end{document}